\documentclass[11pt]{article}

\usepackage[USenglish]{babel} %francais, polish, spanish, ...
\usepackage[T1]{fontenc}

\usepackage{lmodern} %Type1-font for non-english texts and characters
\usepackage{setspace}
\usepackage{bookmark}
\usepackage{amsmath}
\usepackage{amsthm}
\usepackage{amsfonts}
\usepackage{fancyhdr}
\usepackage{graphics}
\usepackage{epsfig}
\usepackage{graphicx}
\usepackage{textcomp}
\usepackage{amsmath}
\usepackage{amsfonts}
\usepackage{amssymb}
\usepackage{mathrsfs}
\usepackage{enumerate} 
\usepackage{epstopdf} 
\usepackage{float}
\usepackage{subcaption}

\usepackage{cleveref}
\usepackage{placeins} 
\usepackage{flafter}  

\usepackage{xcolor}

\numberwithin{equation}{section}

\begin{document}
\begin{titlepage}

\title{Analog Black Holes and Energy Extraction by Super-Radiance from Bose Einstein Condensates (BEC) with Constant Density }
\author{ Bet\"ul Demirkaya\footnote{bdemirkaya@ku.edu.tr}, Tekin Dereli\footnote{tdereli@ku.edu.tr}, Kaan G\"uven\footnote{kguven@ku.edu.tr},  \\ {\small Department of Physics, Ko\c{c} University, 34450 Sar{\i}yer, \.{I}stanbul, Turkey }}

\date{\today}
\maketitle
\begin{abstract}
\noindent 
This paper presents a numerical study of the acoustic superradiance from the single vortex state of a Bose-Einstein condensate (BEC). The draining bathtub model of an incompressible barotropic fluid is adopted to describe the vortex. The propagation of the velocity potential fluctuations are governed by the massless scalar Klein-Gordon wave equation, which establishes the rotating black-hole analogy. Hence, the amplified scattering of these fluctuations from the vortex comprise the superradiance effect. Particular to this study, a coordinate transformation is applied which enables the identification of the event horizon and the ergosphere termwise in the metric. Thus, the respective spectral solutions can be obtained asymptotically at either boundary.  Further, the time-domain calculations of the energy of the propagating perturbations and the independently performed reflection coefficient calculations from the asymptotic solutions of the propagating perturbations are shown to be in very good agreement. While the former solution provides the full dynamical behavior of the superradiance, the latter method gives the frequency spectrum of the superradiance as a function of the rotational frequency of the vortex. Hence, a comprehensive analysis of the superradiance effect can be conducted within this workframe.
\end{abstract}
\vskip 2cm
\end{titlepage}

\maketitle			
\clearpage 

\section{Introduction}
Analogies in physics enable us to observe a particular phenomenon with the same characteristic features in different systems pertaining to disparate mechanisms and space-time-energy scales. The present study takes on such an analogy between a black hole and a vortex state of a Bose-Einstein condensate and focuses on the Hawking radiation, the superradiance of light from black holes, in the form of an acoustic superradiance of sound from a vortex. The analogy initiated by Unruh's calculations showed the equivalence between the background solution of velocity perturbations on a perfect barotropic, irrotational Newtonian fluid and the Klein-Gordon field propagating in a 4-dimensional pseudo-Riemannian manifold, in which the speed of sound plays the role of speed of light \cite{Unruh1981},\cite{barcelo2011analogue}.

The superradiance phenomenon is the amplification of the waves scattered from a black-hole in the presence of ergoregion and it is characterized by a reflection coefficient larger than unity \cite{Penrose1969}, \cite{zel1971generation}. Because this phenomenon occurs in the space-time background of rotating black holes,the analogy could be set for a rotating acoustic black-hole in a liquid \cite{Cardoso2016}, \cite{benone2016superradiance}. The theoretical and computational investigation of the superradiance in various analogous systems have  been  reported by a number of studies. Basak and Majumdar introduced a DBT model of a water vortex and studied the conditions under which the density fluctuations of the fluid exhibit amplified scattering from the water vortex \cite{basak2003superresonance},\cite{dolan2012resonances}. The phenomena is also investigated widely for optical systems \cite{vocke2018rotating},\cite{PhysRevA.86.063821},\cite{philbin2008fiber},\cite{ghazanfari2014acoustic}, relativistic fluids \cite{giacomelli2017rotating}, shallow water systems \cite{richartz2015rotating} .

On the other hand, the experimental studies emerged only within the last few years. Experimental realizations of horizons were reported in
water channels \cite{PhysRevLett.106.021302}, atomic Bose-Einstein condensates (BECs)
\cite{PhysRevLett.105.240401}. Recently, rotational superradiant scattering in a water vortex flow is reported \cite{torres2017rotational}. Acoustic black hole in in a
needle-shaped BEC of 87Rb  is achieved and recently spontaneous Hawking radiation, stimulated by quantum vacuum fluctuations, emanating from an analogue black hole in an atomic Bose-Einstein
condensate is reported \cite{steinhauer2014observation},\cite{steinhauer2016observation}.
  
Motivated by the recent experimental progress in BEC systems, we present here a consolidating study of the temporal and spectral features of the scattering process from a BEC vortex with constant background density. We primarily adopt the draining bathtub model(DBT) introduced by Visser \cite{barcelo2001analogue}, to describe the acoustic black-hole. The time domain solutions are obtained by solving the Klein-Gordon equation for the propagation of acoustic waves, whereas the spectral analysis of the superradiance is conducted by asymptotic solutions of the waves at the event horizon and the ergosphere. The time-domain solutions are obtained by implementing the numerical techniques described mainly in Ref.s \cite{cherubini2005}, \cite{num1}.  In particular, our study demonstrates a very good agreement between the full time-domain and asymptotic frequency domain solutions and reveals spectral features to understand the dependence of superradiance to the rotational speed and to the frequency of the incident waves.
The paper is organized as follows: Section \ref{Theoretical Model} describes briefly the BEC system and gives a theoretical formulation leading to the main (Klein-Gordon) equation. Section \ref{Numerical Model} and \ref{Numerical Results} are devoted to the implementation and computation of the time-domain solutions. Section \ref{Freq Domain} presents the asymptotic solutions in the frequency domain. The last section discusses the main results and concludes the paper.

\section{Theoretical Model}
\label{Theoretical Model}

We begin by a brief description of the Bose-Einstein condensate as the physical system of interest. A quantum system of N interacting bosons in which most of the bosons occupy the same single particle
quantum state, the system can be described by a Hamiltonian of the form;
\begin{eqnarray}
H= \int dx \hat{\Psi}^\dagger(t,x)\left[-\frac{\hbar^2}{2m}\nabla^2 + V_{ext}(x)\right]\hat{\Psi}(t,x) \nonumber \\
 +\frac{1}{2}\int dx dx^\prime \hat{\Psi}^\dagger(t,x) \hat{\Psi}^\dagger(t,x^\prime) V(x-x^\prime)\hat{\Psi}(t,x^\prime) \hat{\Psi}(t,x).
\end{eqnarray}
Here $V_{ext}$ is an external potential and $V(x-x^\prime)$ is the interatomic two-body potential, m is the mass of the bosons and $\hat{\Psi}^\dagger(t,x)$ is the boson field operator which includes the classical contribution $\psi(t,x)$ plus excitations $\hat{\varphi}$.

In the non relativistic limit most of the atoms lie on the ground state and the interatomic interaction is taken as  $V(x-x')=U_0\delta(x-x_0)$, $U_0=4a\pi\hbar^2/m$, where the constant a is called the scattering length. Closed-form equation for weakly interacting bosons, with the potential defined above leads to the time dependent Gross-Pitaevskii(GP) equation:
\begin{equation}
i\hbar\frac{\partial\psi}{\partial t}=\left(-\frac{\hbar^2}{2m}\nabla^2+ V_{ext} +U_0 \left| \psi \right|^2\right)\psi(r,t).
\label{GP}
\end{equation}
Here in hydrodynamic form the wave function can be written in terms of its magnitude and phase:
 \begin{equation}
 \psi(r,t)= \sqrt{\rho}e^{iS}.
 \label{eq:hydro form}
 \end{equation}
Then the density of particles is given by $\rho(t,r)=\left|\psi(t,r)\right|^2$ and the background fluid velocity is defined as $\vec \upsilon=(\hbar/m)\nabla S$. A general review on BEC analogy can be found in \cite{fetter2009rotating}, \cite{macher2009black}. 

 Fluid velocity for the DBT model is defined to have a tangential and radial components,
\begin{equation}
 \vec{ \upsilon}=\upsilon_{\hat{\phi}}+\upsilon _{\hat r}=\frac{-A}{r}\hat{r}+\frac{B}{r}\hat{\phi}\\ 
\label{fluidvel}
 \end{equation}
  where $A$ and $B$ are constants to be determined.

By linearizing the GP equation around some background  $\rho= \rho_0 + \rho_1$ and $S= S_0 + S_1$, we reach two equations defining the density fluctuations and the phase fluctuations:
 \begin{equation}
            \frac{{\partial \rho_1 }}{{\partial t}} + \frac{\hbar}{m}\nabla  \cdot (\rho_0 \nabla S_1) + \nabla  \cdot (\rho_1 \upsilon)= 0,
						 \label{densfl}
  \end{equation}
  \begin{equation}
    \partial_t S_1 =-\upsilon \cdot \nabla S_1 - \frac{U_0}{\hbar}\rho_1 + \frac{\hbar^2}{2m}D_2 \rho_1,
            \end{equation}
						where $D_2$ is given by
\begin{equation}
D_2=\frac{1}{2\sqrt{\rho_0}}\nabla^2 \frac{\rho_1}{\sqrt{\rho_0}}-\frac{\rho_1}{2\rho^{3/2}_{0}}\nabla^2\sqrt{\rho_0}.
\label{qunapres}
\end{equation}
Hydrodynamic approximation (quasiclassical approximation) where $D_2 = 0$ is justified by pointing out that $D_2$ is relatively small compared to other terms. The pressure term is of the order $U_0 \rho/R$ while the quantum pressure term is of the order $\hbar^2 / m R^3$, where $R$ is the spatial scale \cite{pethick2002bose}. This implies that for 
\begin{equation}
R>> \frac{\hbar}{\sqrt{2mU_0 \rho}}
\label{eq:healing}
\end{equation}
hydrodynamic approximation hold, that gives the healing length parameter. Therefore, the approximation leading to the KG equation is given in its final form as

 \begin{equation}
\frac{\partial}{\partial t}\left[\frac{\rho_0}{c^2}\left(\frac{\partial S_1}{\partial t} + \vec \upsilon \cdot \nabla S_1\right]\right) - \nabla \cdot \left(\rho_0 \nabla S_1\right) + \nabla \cdot \left[\frac{\rho_0}{c^2}\left(\frac{\partial S_1}{\partial t} + \vec \upsilon \cdot \nabla S_1\right)\right]=0
\label{pdeKG}
 \end{equation}
where the speed of sound is defined by $c=\sqrt{\rho U_0/m}$.
The stationary, axially symmetric metric associated with this configuration will be;

\begin{equation}
ds^2 = \frac{\rho_0}{c}\left[-(c^2-\frac{A^2+B^2}{r^2})dt^2 + \frac{2A}{r}dtdr - 2Bdtd\phi + dr^2 + r^2d\phi^2 + dz^2\right].
\label{line}
\end{equation}
For a constant density profile, for which the speed of sound is constant, the resulting equation is the massless Klein-Gordon wave equation for linear perturbations of the velocity potential, or phase of the wave function.
	
	\subsection{Coordinate Transformations}

The coordinate transformation indicated below is particularly useful to  minimize the number of off-diagonal components of the metric, leaving only one, which helps in analyzing the asymptotic behavior, and to reveal the event horizon and the ergosphere.
	  \begin{align}
                dt=dt^{*}-g*dr && d\phi=d\phi^{*} - h*dr && r=r^{*} && z=z^{*},
             \end{align}
where $h=-(AB)/(r(A^2 - c^2r^2))$ and $g=-(Ar)/(A^2 - c^2r^2)$. We drop the *-superscript in the following part of the formulation. The line equation takes the form

\begin{equation}
ds^2=\frac{\rho_0}{c}\left[-\left(1-\frac{A^2+B^2}{c^2r^2}\right)dt^2 + \left(1-\frac{A^2}{c^2r^2}\right)^{-1}dr^{2}-\frac{2Bd\phi dt}{c} + r^2d\phi^2 + dz^2 \right].
\label{eq:newline}
\end{equation}
Now it is easy to see the distinction between the event horizon and the ergosphere. From the definitions, as for the Kerr black hole in general relativity, the radius of the ergosphere is given by the vanishing of $g_{00}$ and the coordinate singularity of the metric signifies the event horizon. For the DBT model, they read as
\begin{align}
 r_{event}= A/c, && r_{ergo} = (A^2 + B^2)^{1/2}/c> r_{event}.
\end{align}

\section{Numerical Model In The Time Domain}
\label{Numerical Model}

In order to solve the Eq.\ref{pdeKG}, first we write the line element in the form;
\begin{equation}
ds^2= - \alpha^2dt^2 + \gamma_{ij}(dx^i + \beta^idt)(dx^j + \beta^jdt),
\end{equation}
where $\alpha=c$, $\gamma_{ij}= diag(1,r^2,1)$ and $\beta^i = (A/r, - B/r^2, 0)$. For numerical integration of the scalar field perturbations $\Psi = S_1(r,t)$, we introduce two conjugate fields;
\begin{align}
\Phi = \frac{\partial {\Psi}}{\partial x^i} && {\Pi}=-\frac{1}{\alpha}\left(\frac{\partial {\Psi}}{\partial t}  - \beta^i {\Phi}_i\right),
\label{conjugate}
\end{align}
where  ${\Psi} = \psi_1 (t,r)e^{im\phi}e^{ikz}$, ${\Pi} = \pi_1 (t,r)e^{im\phi}e^{ikz}$ and ${\Phi} = \phi_1 (t,r)e^{im\phi}e^{ikz}$ and $(k,m)$ are the axial and azimuthal wave numbers \cite{num1}, \cite{Andersson1998}. In this work, in accordance with the BEC vortex stability conditions, we consider the azimuthal wave numbers of m = 0 and 1 only. For m = 2, the vortex would decay into two separate vortices of m =1  \cite{desyatnikov2005optical}, \cite{zhang2007numerical}. Then our hyperbolic system reads,
\begin{align}
&\partial_t \pi_1 + c\partial_r \phi_1 - \frac{A}{r}\partial_r \pi_1 = -imB\pi_1/r^2 + c(k^2 + m^2/r^2)\psi_1 -c\phi_1/r\nonumber\\
&\partial_t \psi_1  - \frac{A}{r}\partial_r \psi_1 = -imB\psi_1/r^2 -c\pi_1\nonumber\\
&\partial_t \phi_1 + c\partial_r \pi_1 - \frac{A}{r}\partial_r \phi_1 = 2imB\psi_1/r^3 - (A+imB)\phi_1/r^2.
\label{setofpart}
\end{align}
The remaining first order set of coupled PDEs are much easier to handle than the hyperbolic PDE above that we start with. 

\section{Numerical Results}
\label{Numerical Results}
The numerical challenges of using a constrained evolution scheme is mostly about avoiding constraint violations and other possible numerical issues which may be associated to solver type and settings, element type and size, meshing, tolerances, etc. All of these ingredients must be fine tuned in the computation to get proper results. However, we still have the freedom to try different interior boundary conditions because excision, i.e by placing the boundary inside the horizon and excises its interior from the computational domain. In theory at least, nothing physical inside the black hole can influence any of the physics outside the horizon \cite{excision}. 

This main section is organized as follows: We first calculate the time evolution of the perturbations of the velocity potential by solving the equations \ref{setofpart} and the energy of the perturbations given further in Eq.\ref{eq:energy}. Based on the ranges of the model parameters $\Omega$ and $\omega$, superradiant and superradiant cases are demonstrated for comparison.

The initial value is chosen as a Gaussian pulse centered at $r_0$, modulated by a monochromatic wave \cite{Andersson1998}:
\begin{equation}
\psi_1(0,r)=Aexp\left[-(r-r_0+ct)^2/b^2 - i\omega(r-r_0 +ct)/c\right].
\label{initial}
\end{equation}

\begin{figure}[H]
\centering
		\includegraphics[width=1\textwidth]{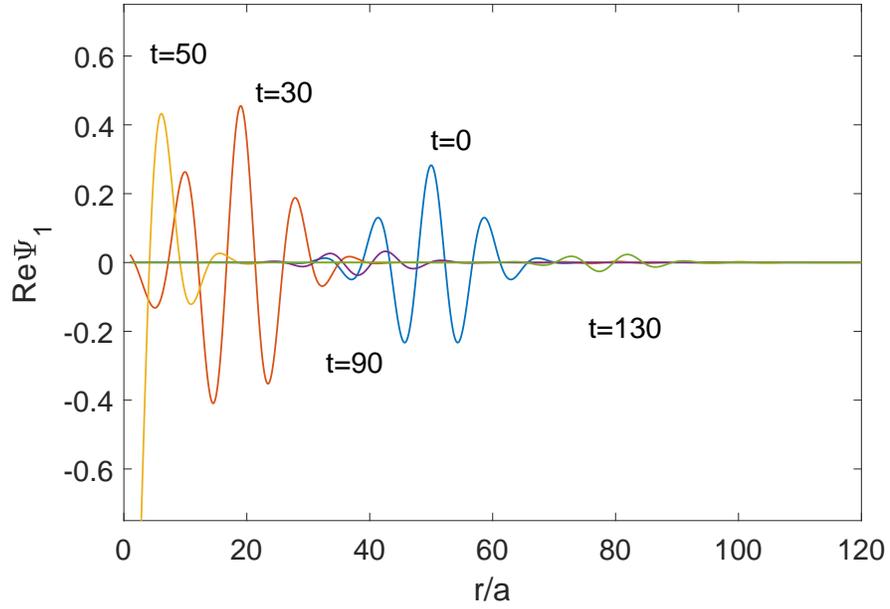}
		\caption{Snapshots of the time evolution of the perturbation, for the case m=0 as a function of distance $r$ from the vortex, for $r_0$=50a and $b$=10a with $\omega$=0.7c/a, $\Omega$=1.4c/a  }
	\label{fig:m0casewave10}
\end{figure}

\begin{figure}[H]
	\centering
		\includegraphics[width=1\textwidth]{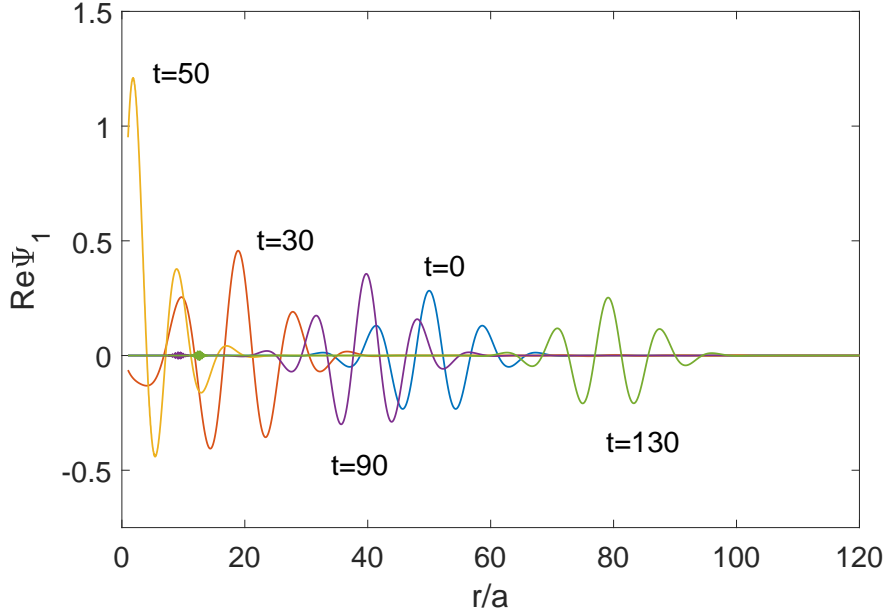}
		\caption{Snapshots of the time  evolution of the perturbation, for the case m=1 as a function of distance $r$ from the vortex. The parameters used are in Fig. \ref{fig:m0casewave10} }
	\label{fig:m1casewave10}
\end{figure}

The equation system \ref{setofpart} is integrated numerically using Matlab PDE solver by modifying equation format and boundary conditions accordingly \cite{cherubini2005}. The set of equations \ref{setofpart} allows us to decompose $\Pi,\Phi$ and $\psi$ into characteristic fields that propagate along null ray
\begin{align} 
u^+ \propto \Pi + \Phi && u^- \propto \Pi - \Phi
\end{align}
At large distances purely outgoing wave is implemented such that $u^- =0 $, $\pi_1 = \phi_1$. While outer boundary condition has to be well-behaved for the calculation, simulation  ends before the wave reaches the outer boundary, therefore nature of the reflections
produced when a wave passes through the boundary is largely irrelevant. And for the inner boundary, because of the excision, no boundary condition is set. The computational spatial (radial) and time domain are set as $0.2 < r < 150, 0 < t < 150$, with discretization steps of $\Delta r = 0.05, \Delta t = 0.05$, respectively. Inner boundary is set according to the constraint values at the event horizon. While the waves are free to propagate inside the horizon, reflected waves from the $r=0$ should not effect the expected result, such that inner boundary should not be too close to the singularity $r=0$ or to the event horizon $r=1a$.
 The domain is sufficiently large to achieve steady state solutions, whereas the discretization steps provide good accuracy for the solution and for the constraint equations.

The incident wave is a cylindrically imploding Gaussian wave, centered at $r_0 = 50a$ with a width of $b = 10a$ and azimuthal wavenumber $k = 0.02/a$. Here, c is the propagation speed of sound in the condensate and a is location of the event horizon. Both parameters are scaled to unity. We note that the location of the incident wave should be chosen numerically far enough so that the scattering outcome is independent from the location of the incident wave. The angular speed of the vortex is $\Omega$. In the present calculations, we consider values of $\Omega$ up to $\Omega = 4c/a$. The frequency of the incident wave is $\omega = \Omega/2$.

Fig.\ref{fig:m0casewave10} and Fig.\ref{fig:m1casewave10} show the snapshots of the time evolution of the initial Gaussian wave. While the wave is getting closer to the horizon, it is affected by the potential  and gets reflected. As expected, the perturbation for the non-superradiant ($m=0$) case goes to zero while in the superradiant case (m=1) it gets amplified through backscattering.

\begin{figure}[H]
  \begin{subfigure}[t]{.45\textwidth}
    \centering
    \includegraphics[width=\linewidth]{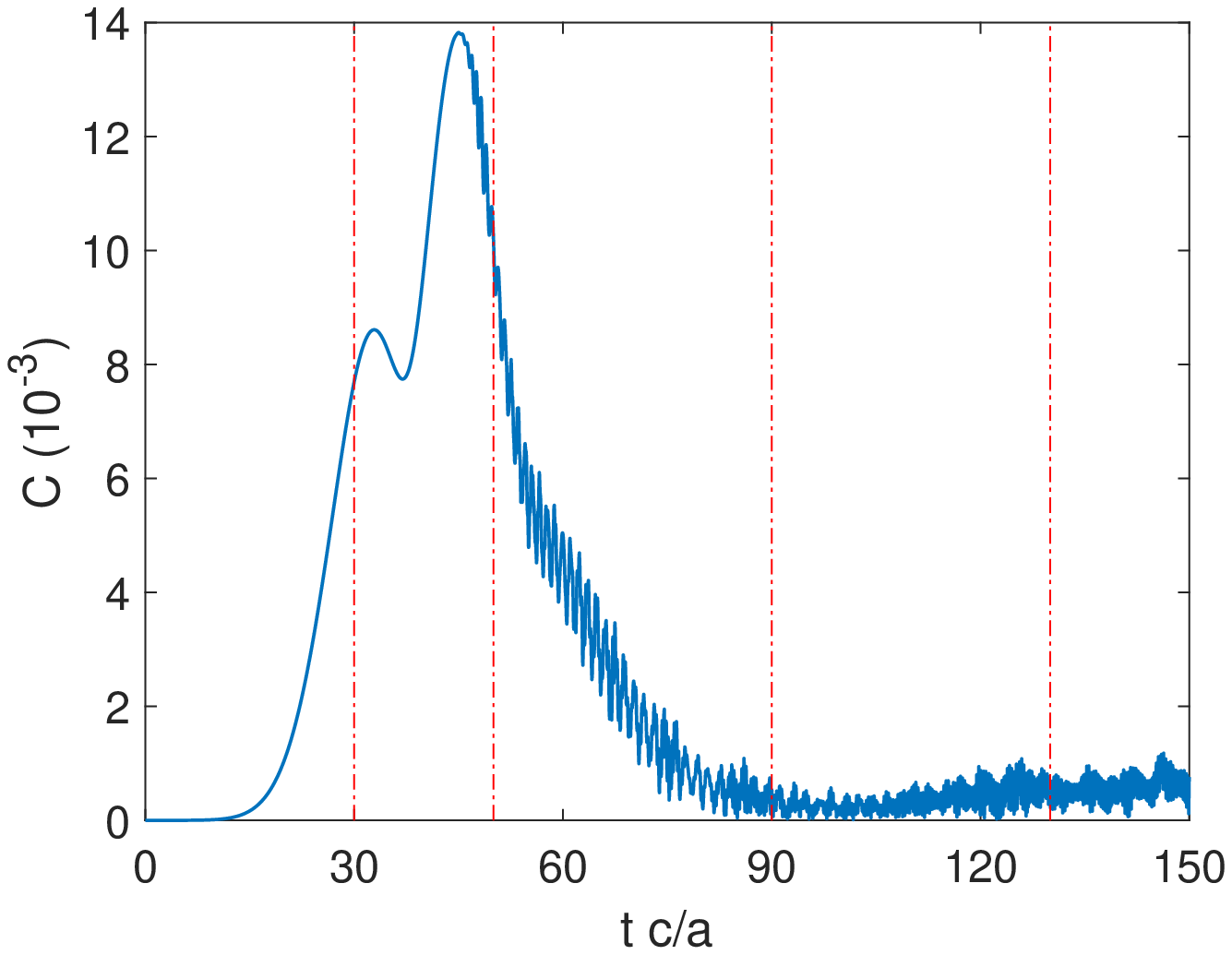}
  \end{subfigure}
   %%\hfill
  \begin{subfigure}[t]{.45\textwidth}
    \centering
    \includegraphics[width=\linewidth]{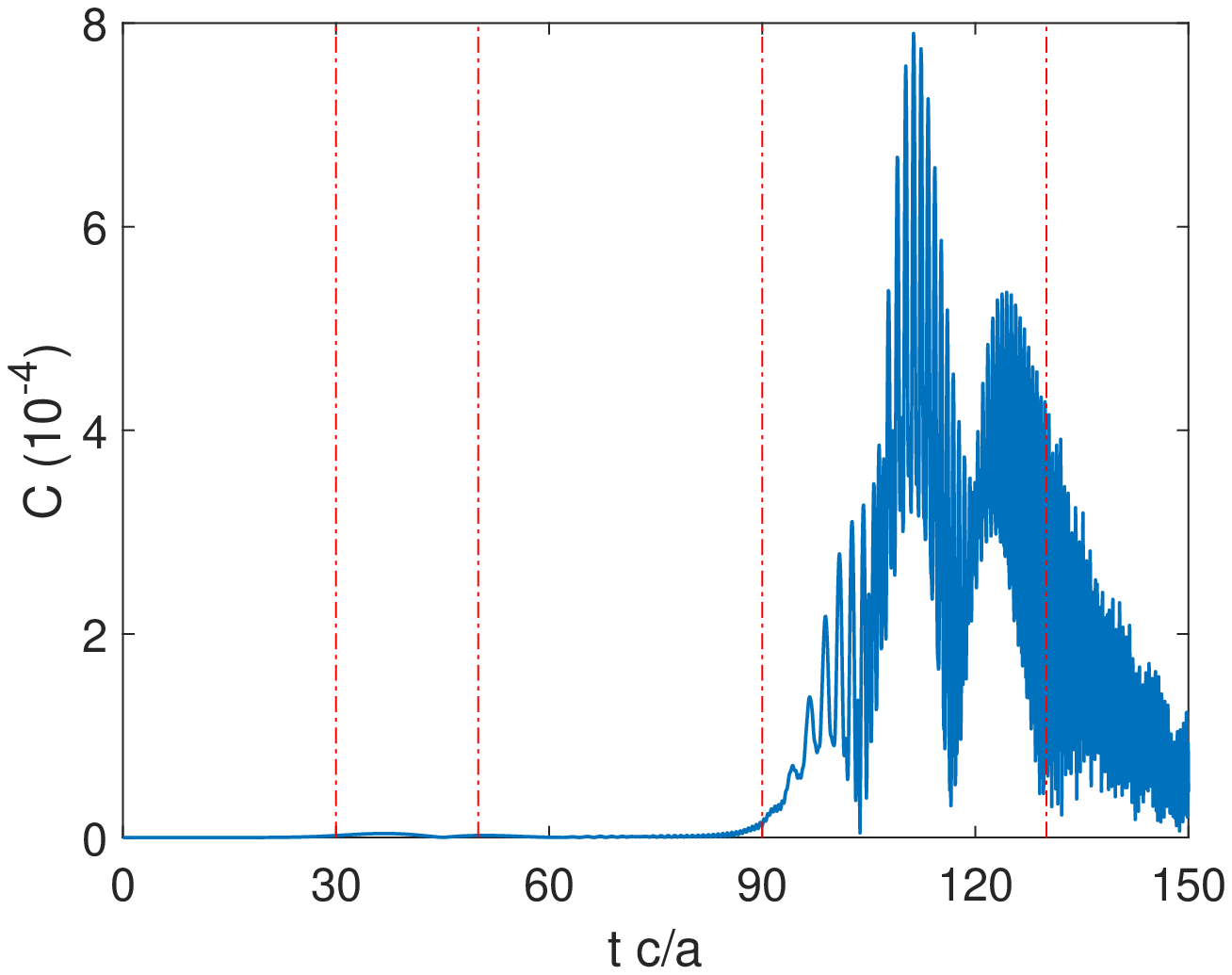}
  \end{subfigure}
	 %%\hfill
	\caption{Constraint violations at event horizon(a) and outer boundary($r=80a$)(b) for superradiance (m=1). The parameters used are in Fig. \ref{fig:m0casewave10}. Dotted lines signify the time frames(t=30,50,90,130) shown in Fig\ref{fig:m0casewave10} and Fig.\ref{fig:m1casewave10}}
	\label{Constraints_srad}
\end{figure}
\begin{figure}[H]
  \begin{subfigure}[t]{.45\textwidth}
    \centering
    \includegraphics[width=\linewidth]{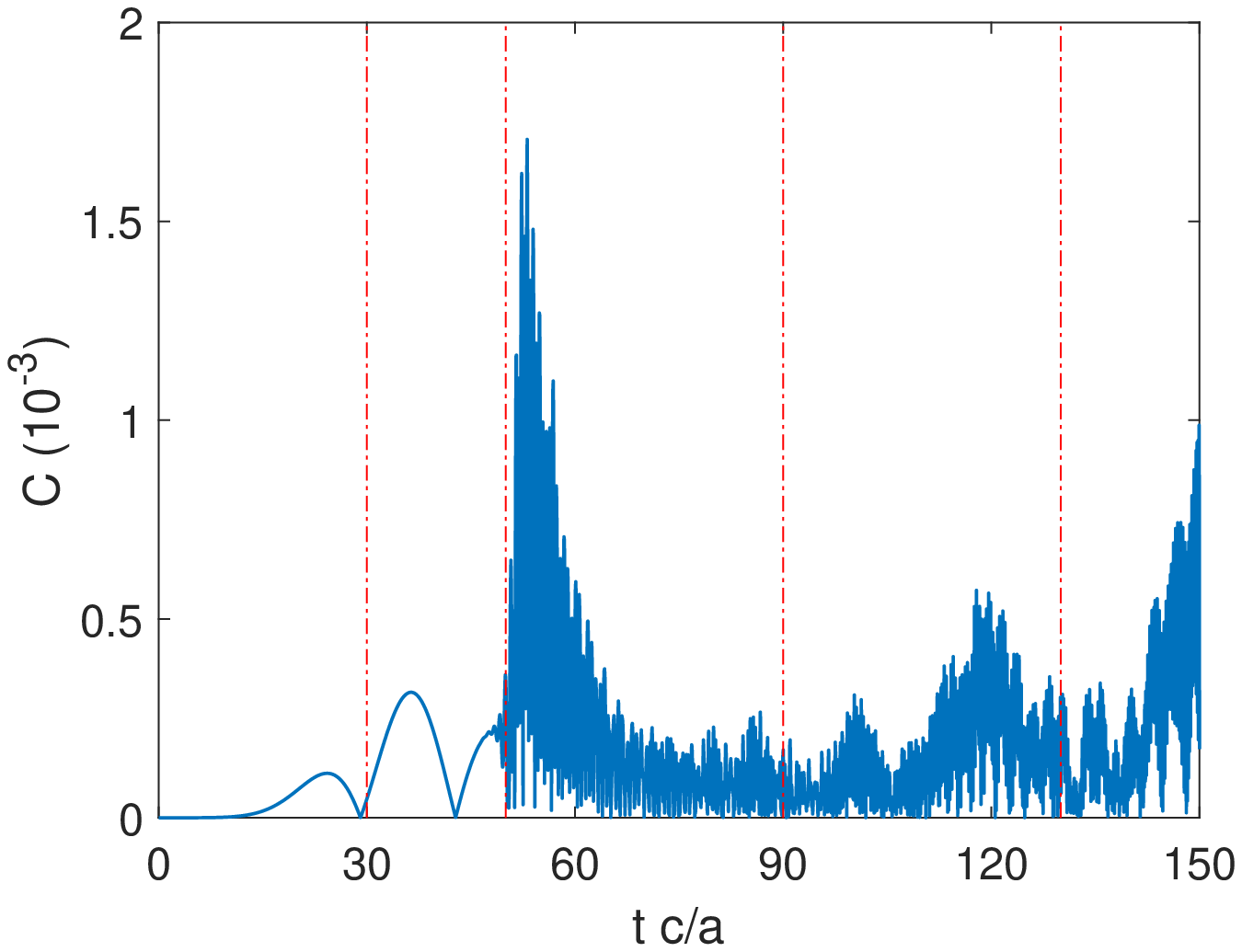}
  \end{subfigure}
 %%\hfill
  \begin{subfigure}[t]{.45\textwidth}
    \centering
    \includegraphics[width=\linewidth]{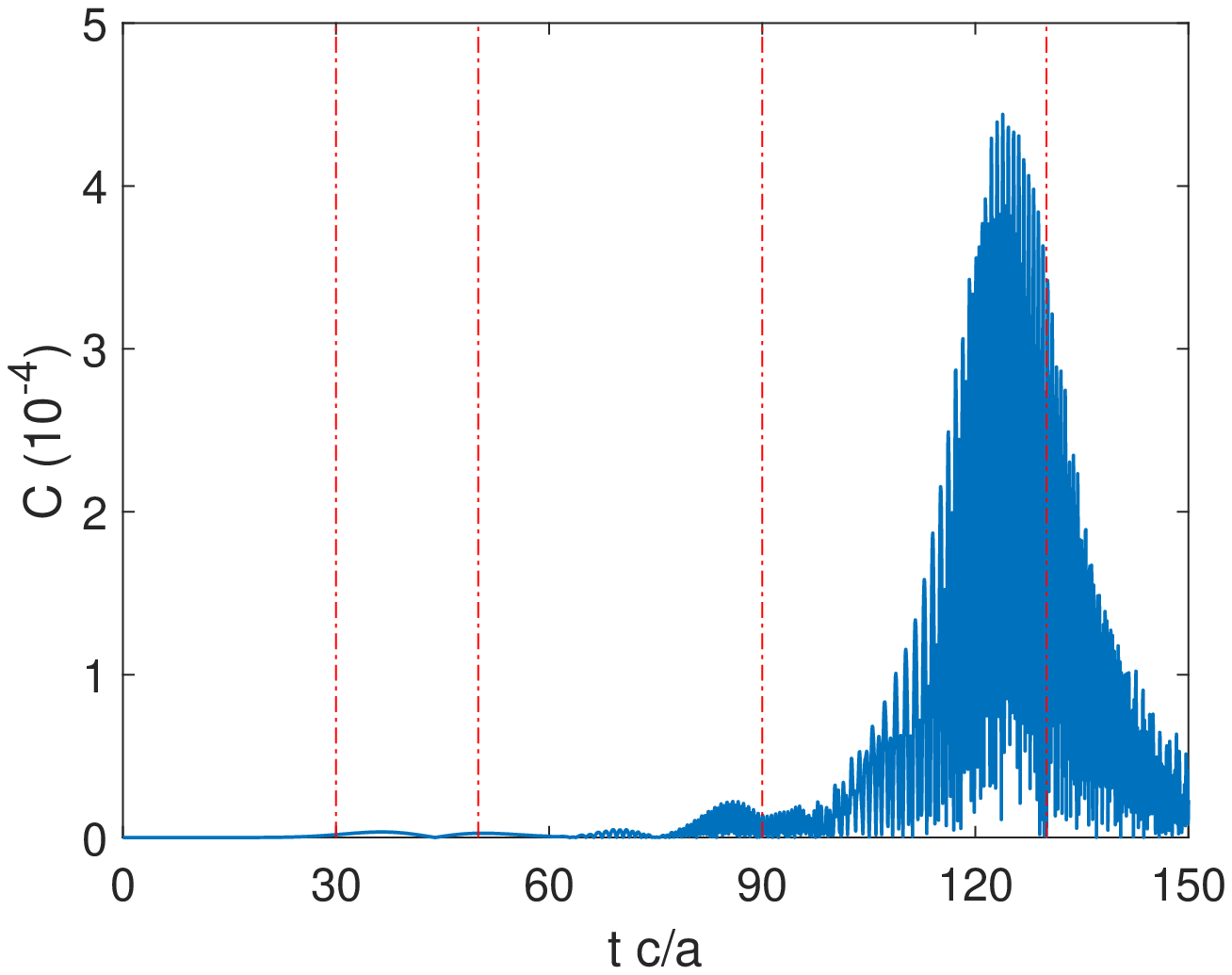}
  \end{subfigure}
	 %%\hfill
	\caption{Constraint violations at event horizon(a) and outer boundary($r=80a$)(b) for non-superradiance (m=0). The parameters used are in Fig. \ref{fig:m0casewave10}}
	\label{Constraints_nonsrad}
\end{figure}

In order to check the quality of the numerical analysis, we monitor the constraint value C, from the definition of $\Phi$, Eq.\ref{conjugate} 
 
\begin{equation}
C = \left|\partial_r \psi_1 - \phi_1\right| .
\label{eq:constraint}
\end{equation} 
The constraint values should be closer to zero and not increase in time such that any unphysical waves, backscattered radiation would not overpower the actual results. 
Since the reflected wave can only reach approximately to $r=80/a$ at the time $t=130c/a$, as seen in the Fig\ref{fig:m0casewave10} and Fig.\ref{fig:m1casewave10}, $r=80/a$  point is chosen to calculate the constraint violations for the outer boundary. Even though the outer boundary for the simulation is at $r=150/a$,  where the wave can not reach during the computation time.

It shows from the Fig.\ref{Constraints_nonsrad} and Fig.\ref{Constraints_srad} that the constrained value, C, does not grow indefinitely in time and remains under a certain value. At the inner horizon, scaled to $1a$, constraint values shown to be larger than outer boundary,reaches the maximum value around $t=50c/a$, when the wave is partially absorbed by the event horizon. But still remains small enough that the violations are negligible.  In addition, we observe that for larger frequencies, we have to keep an eye on the inner constraint violations more closely  to check that  results are meaningful, since the simulations become unstable much faster. 

The time variation of the energy of wave packet is given by 
\begin{equation}
E(t)=(\rho \hbar^2/2M)\int^{2 \pi}_{0}d \phi \int^{H}_{0}dz \int^{r_{max}}_{1} (\nabla \psi_1)^2 r dr.
\label{eq:energy}
\end{equation}

\begin{figure}[H]
	\centering
		\includegraphics[width=0.8\textwidth]{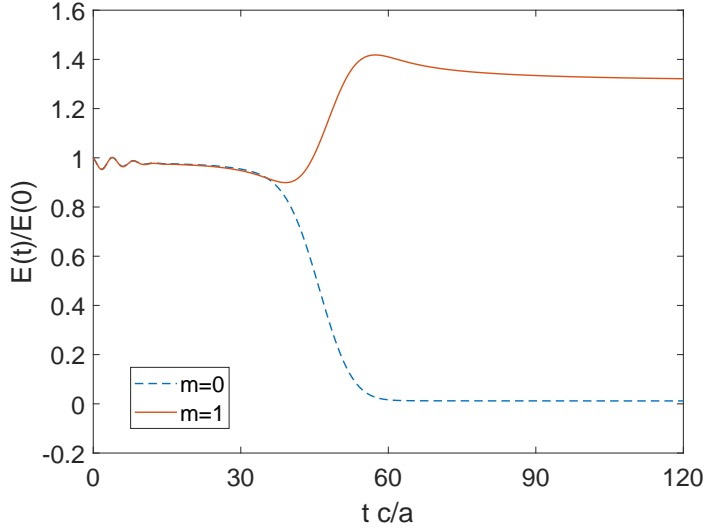}
		\caption{Time evolution of the energy gain of the wave packet, superradiant m=1 case and non-superradiant m=0 case. The parameters used are in Fig. \ref{fig:m0casewave10}}
	\label{fig:superradiance}
\end{figure}
Figure \ref{fig:superradiance} shows the time evolution of the energy of the wave, normalized by the energy of the incident wave, for the  non-radiant (dashed blue curve) and for the superradiant (solid red curve) cases respectively. Note that the wave arrives to the event horizon near t = 35 c/a. In the non-radiant case, all the impinging energy is lost to the vortex sink. In the superradiant case conditions the scattering process extracts energy from the ergosphere and the energy of the backscattered wave exceeds its incident value. Also, we did not calculate the total energy densities but the energy densities per unit length in z-direction \cite{pethick2002bose}.

\begin{figure}[H]
  \begin{subfigure}[t]{.5\textwidth}
    \centering
    \includegraphics[width=\linewidth]{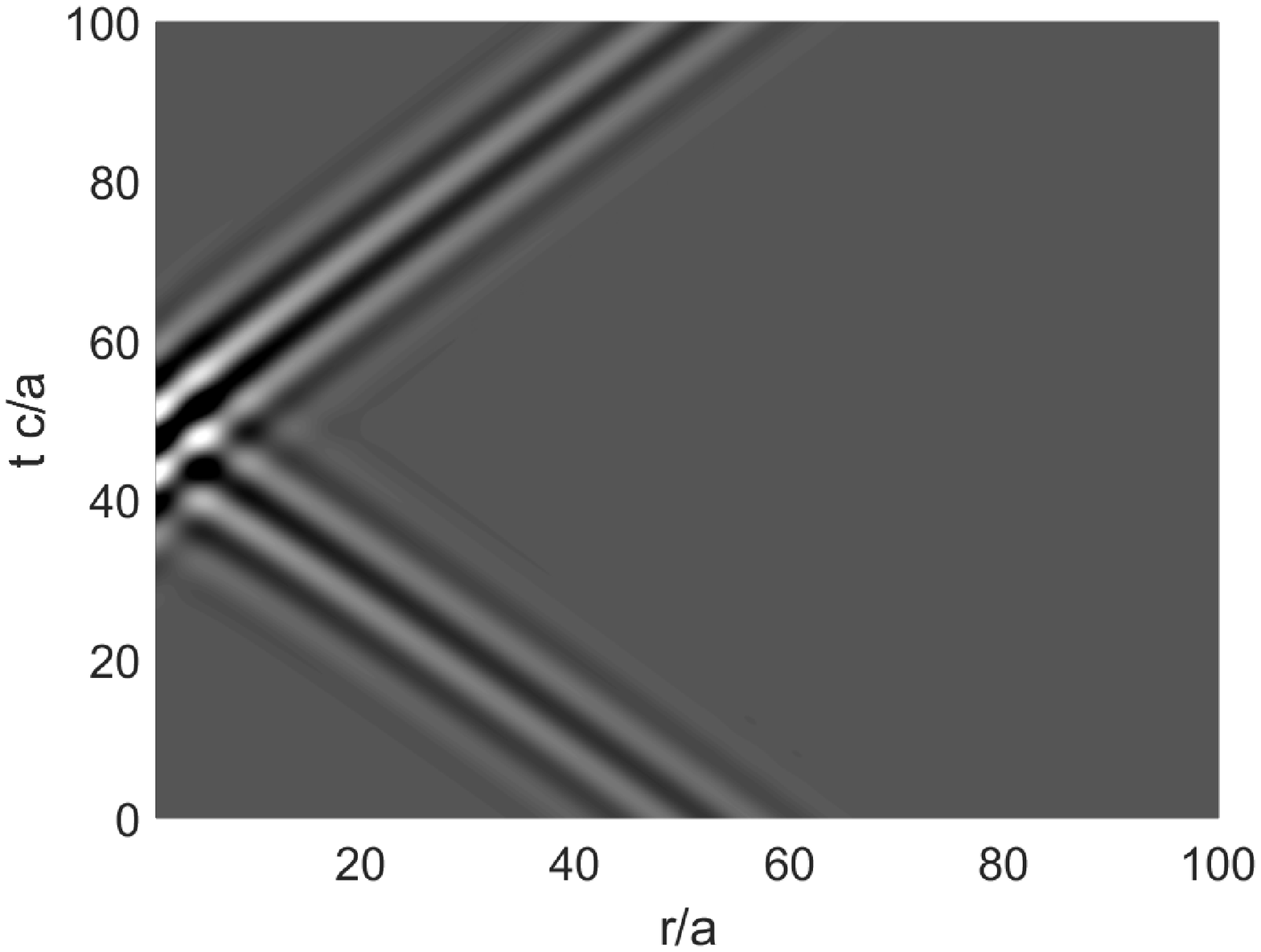}%%{Fig6a.eps}
  \end{subfigure}
  %%\hfill
  \begin{subfigure}[t]{.5\textwidth}
    \centering
    \includegraphics[width=\linewidth]{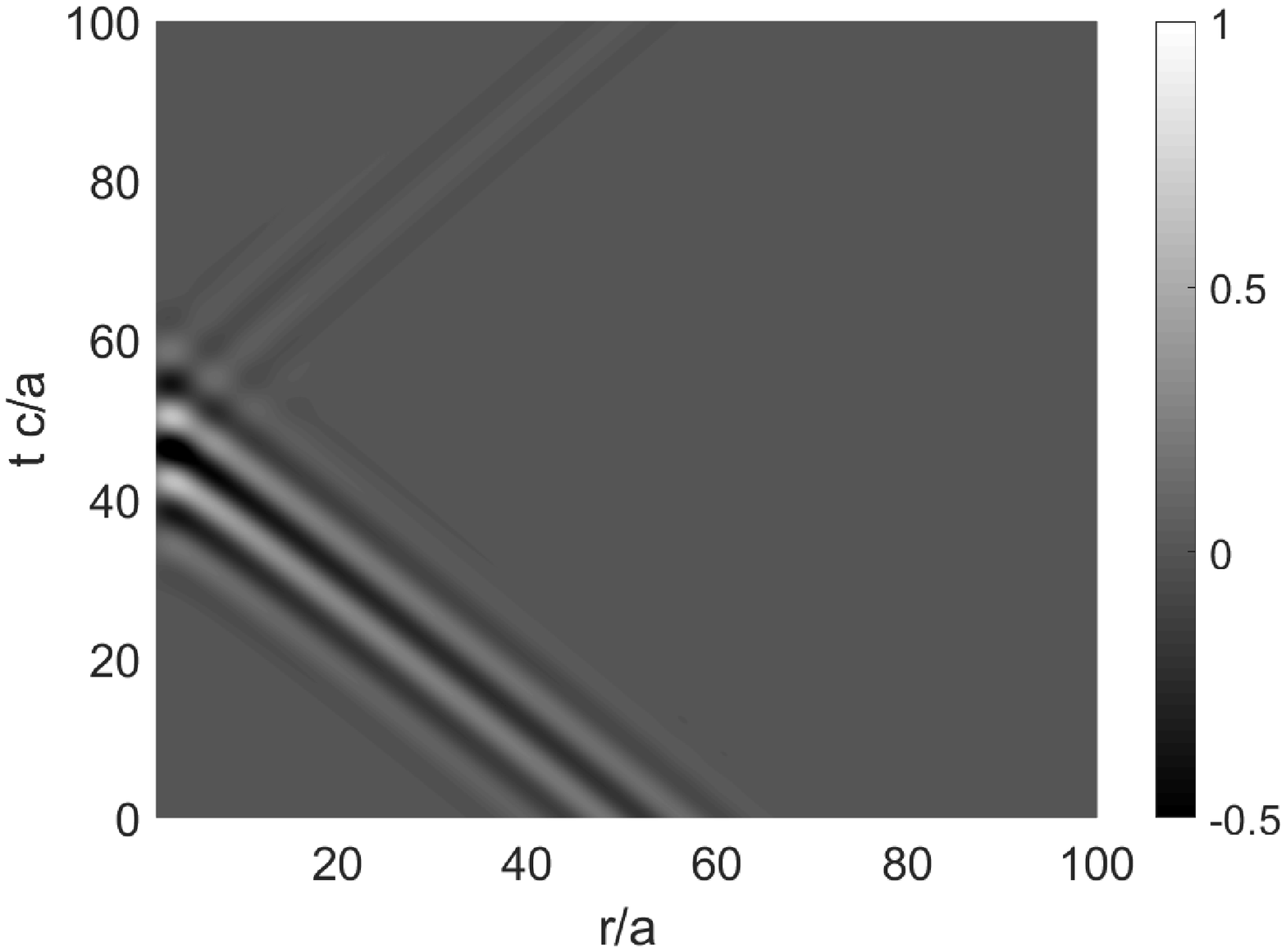}%%{Fig6b.eps}
  \end{subfigure}
	\caption{Density fluctuations, $\rho_1$ in r-t plane for superradiance case, m=1 (left) and m=0 (right).}
	\label{densflfig}
\end{figure}

\begin{figure}[H]
  \begin{subfigure}[t]{.5\textwidth}
    \centering
    \includegraphics[width=\linewidth]{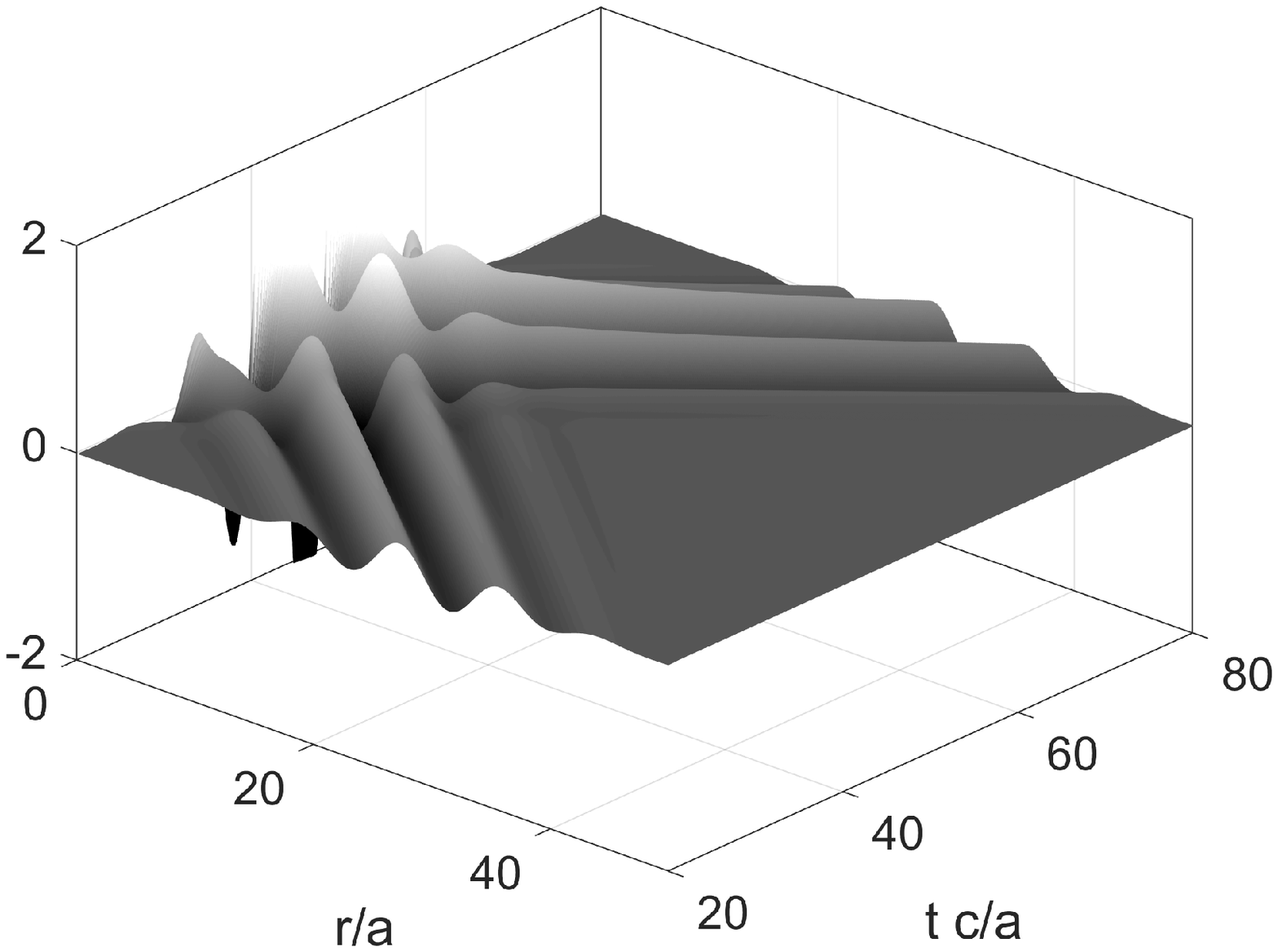}
  \end{subfigure}
   %%\hfill
  \begin{subfigure}[t]{.5\textwidth}
    \centering
    \includegraphics[width=\linewidth]{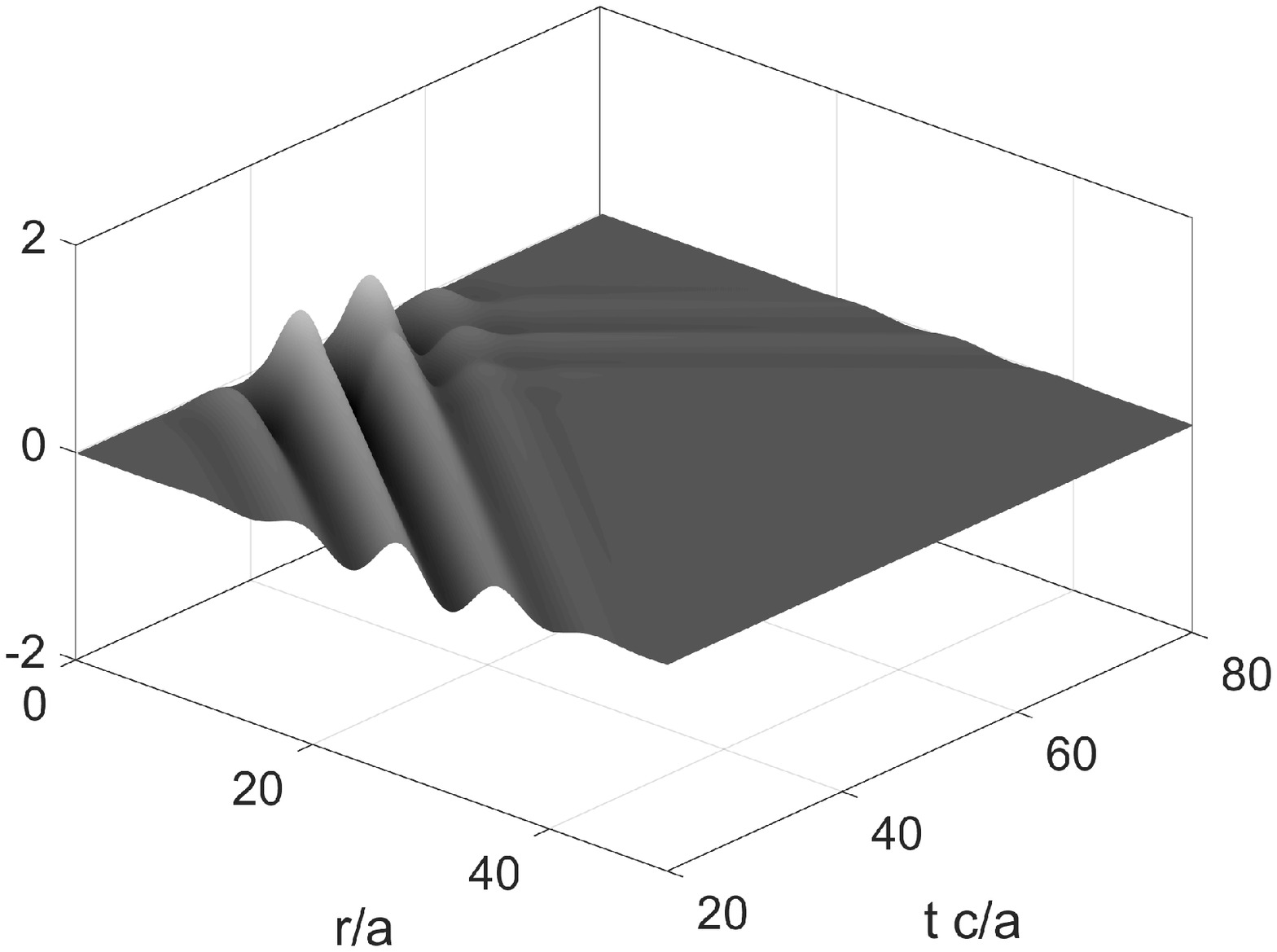}
  \end{subfigure}
	 %%\hfill
	\caption{Closeups of the data in Fig.\ref{densflfig}.}
	\label{densflfigclose}
\end{figure}

\begin{figure}
    \centering
    	\includegraphics[width=0.8\textwidth]{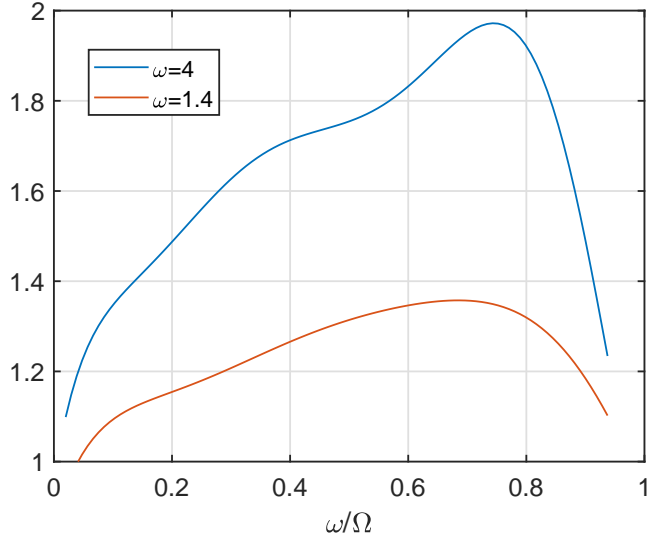}
	\caption{The Energy at $t_{final}=150$ $c/a$ normalized to its initial value $E(t=0)$ as a function of $\omega / \Omega$ where  $0<\omega<\Omega_i$ .The parameters used in the calculations are  $\Omega_i=1.4$, $4$, $r_0 = 50a$, $b=10a$.}
\label{enegyvsomega}
\end{figure}

The time evolution of the density fluctuations associated with the acoustic wave propagation are plotted Fig.\ref{densflfig} for superradiant and non-radiant cases. Figure \ref{densflfigclose} give the detailed view of the propagation of fluctuations near the event-horizon. Sudden increase in the density fluctuations for the superradiance case, shown in Fig. \ref{densflfigclose}, stays inside the event horizon, r=1. 

Evidently, the numerical treatment of the region beyond the event horizon is inherently prone to numerical instabilities. We found that at large time scales after the scattering event, noise fluctuations emerge within the event horizon, which can propagate into the real space and render the simulation results unacceptable. Furthermore, this time scale decreases with increasing omega. Thus, simulation times are adjusted to avoid this problem. Fortunately, the existence of the event horizon allows non-strict boundary conditions so that the numerical instability contained withing the event horizon (i.e. $r<r_{event}$) does not affect the results in the real space domain.

The amplification factor of the reflected wave as a function of the ratio $\omega/\Omega$ is plotted in Fig.\ref{enegyvsomega}, for $\Omega = 1.4$, and $\Omega = 4$, respectively. The amplification increases rather monotonically up to a certain $\omega/\Omega$ ratio, which depends on the particular value of $\Omega$. After that, the amplification decreases rapidly to unity as $\omega/\Omega$ approaches unity. Thus, the maximum superradiance occurs at a particular $\omega$ of the incident wave, in relation to the $\Omega$ of the vortex. In the next section, we will analyze this behavior in the frequency domain.

\section{Numerical Model In The Frequency Domain}
\label{Freq Domain}
In this section we analyze the Klein Gordon equation (Eq.\ref{pdeKG}) in the frequency domain. Using separation of variables, the formal solution of the KG is expressed as

 \begin{equation}
 \label{domain}
 \psi  = {e^{ - i\omega t}}{e^{im\phi }}{e^{ikz}}P(r),
 \end{equation} 
 where $k$ and $m$ are the axial and azimuthal wave numbers, respectively. To avoid polydromy problems \cite{cherubini2005}, that is to make $\psi $ single valued, $m$ should be taken as an integer and $k$ a real number defined by the boundary conditions along the z axis.

 By inserting (\ref{domain}) into (\ref{pdeKG}), we obtain a second order ODE for the radial part:
 \begin{align}
 \frac{{{d ^2}P}}{{d {r^2}}}&
  + \left( \frac{A^2 + r^2c^2 +2iA(Bm-r^2\omega)}{r(r^2c^2-A^2)} \right)\frac{{d P}}{{d r}}\nonumber\\
  \nonumber\\
 &+\left( \frac{2iABm -B^2m^2 + c^2m^2r^2 + 2Bm\omega r^2 -r^4\omega^2  + c^2k^2r^4}{r^2(r^2c^2- A^2)} \right)P = 0.\label{radial}
 \end{align}
We substitute $P=R(r)H(r_*)$ with a Regge-Wheeler tortoise coordinate, $r_*$, which will map $r\in[r_H,\infty]$ to $r_*\in[-\infty,+\infty]$:
      \begin{equation}
      {r_ * } = \int {\frac{{{r^2}}}{{{r^2} - {A^2/c^2}}}dr}.
			\label{toro}
      \end{equation}
Then after a few careful calculations \cite{marques2011acoustic}, the radial equation takes the final form
			 \begin{equation}\frac{{{d ^2}H(r_*)}}{{d {r_*}^2}} + \left( \frac{{{\omega ^2}}}{{{c^2}}} -V(r)  \right)H(r_*) = 0,
			\label{sch}
    \end{equation}
		where
		\begin{equation}
		V= k^2(1-\frac{A^2}{r^2c^2}) - \frac{{5{A^4}}}{{4{c^4r^6}}} - \frac{{A^2\left( {{m^2} - 3/2} \right) + B^2{m^2}}}{{{c^2}{r^4}}} - \frac{1}{{4{r^2}{c^2}}}\left( {{c^2} - 4{m^2}c^2 - 8B\omega } \right).
		\end{equation}
Near the event horizon and at the far field $(r \rightarrow +\infty)$, the asymptotic solutions are given by the harmonic functions, 
					 \begin{equation}
      H({r_*}) = {e^{\frac{i\omega_{+}{r_*}}{c}}} + {{\mathop{\rm Re}\nolimits} ^{ \frac{- i\omega_{+} {r_*}}{c}}} ,  r^* \stackrel{}{\rightarrow} +\infty
			\label{infsol}
   \end{equation}
	
	 \begin{equation}
      H({r_*}) = T{e^{ \frac{- i(\omega  - \Omega m){r_*}}{c}}}, r^* \stackrel{}{\rightarrow} -\infty
      \end{equation}
		where $\omega^2_{+} = \omega^2 - k^2c^2$ and $B=\Omega A^2/c^2$.
		The equality of the Wronskian of these solution at asymptotics gives
		\begin{equation}
             1 - {\left| R \right|^2} = \left( {\frac{{\omega  - m\Omega }}{\omega_+ }} \right)\left| {{T^2}} \right|
						\label{supercond}
     \end{equation}
 where $R$ and $T$  are the amplitudes of the reflection and transmission coefficients of the scattered wave. It shows that when the superresonance condition,$\omega<m\Omega$, is satisfied, reflection coefficient has a magnitude larger than unity \cite{Choy},\cite{Berti2004}.

\begin{figure}[H]
    \centering
    \includegraphics[width=0.8\linewidth]{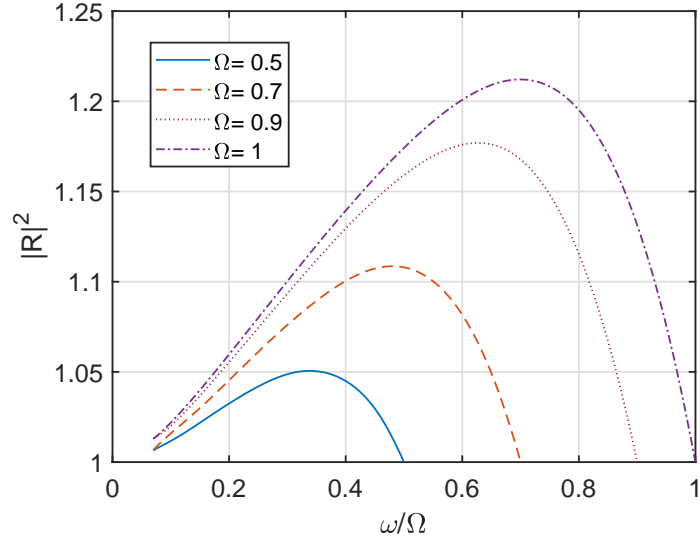}
		\caption{Reflection coefficients as a function of $\omega$, calculated in the range $0<\omega < m\Omega_i$. Parameters are $m=1$, and $\Omega_i$ =0.5, 0.7, 0.9, 1.}
		\label{refcoef1}
\end{figure}
								
\begin{figure}[H]
    \centering
    \includegraphics[width=0.8\linewidth]{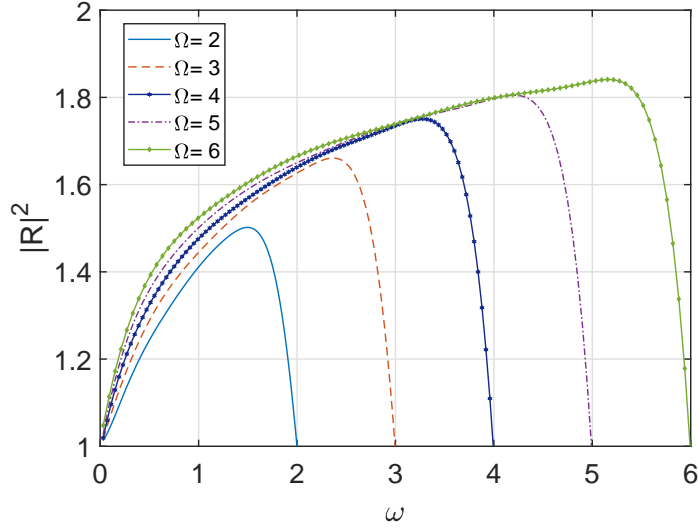}
	\caption{Reflection coefficients as a function of $\omega$, calculated in the range $0<\omega < m\Omega_i$. Parameters are $m=1$, and $\Omega_i$ =2, 3, 4, 5, 6.}
	\label{refcoef2}
	\end{figure}
	Eq.\ref{supercond} reveals the superradiance condition clearly (i.e $\omega < m \Omega$) and gives the full spectral behavior of the reflection coefficient. Thus, we can obtain the reflection coefficient through the Fourier components of the asymptotic far field solution, which is obtained through Eq.\ref{toro} and Eq.\ref{sch}. Figure \ref{refcoef1} and Fig.\ref{refcoef2} show the reflection coefficient as a function of  incident wave frequency for different values of the angular speed of the vortex ($\Omega$) (that is the horizontal axis represents multiple ranges $0<\omega < \Omega_i$). Fig.\ref{refcoef1} is for $\Omega < 1$, Fig.\ref{refcoef2} shows the range $2<\Omega<6$. Here, we used the same model parameters as in the time-domain solution presented in the previous section. 	
					
\begin{figure}[H]
    \centering
    \includegraphics[width=0.8\linewidth]{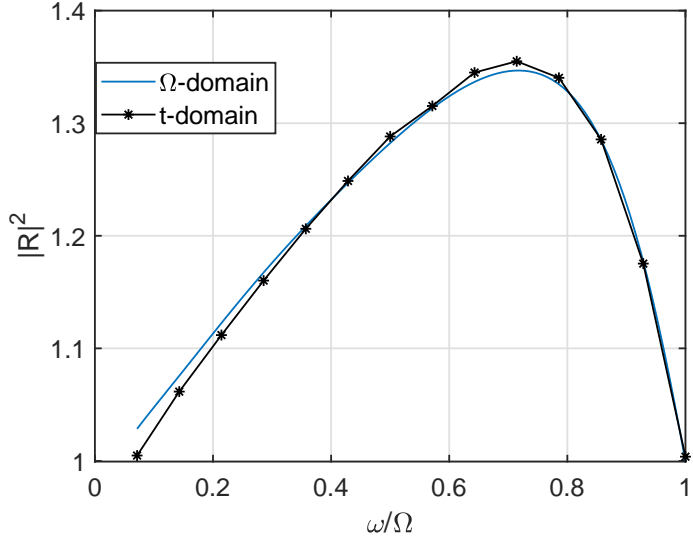}
		\caption{Reflection coefficients of the scattered wave with parameters given under Fig.1  calculated in the time domain and the frequency domain}
		\label{both}
  \end{figure}
 In the frequency domain, coordinate transformation allowed us to carry the calculations outside the event horizon with the asymptotic solutions and reflection coefficient defined in Eq.\ref{infsol} is calculated. 
But, in the time domain, no coordinate transformation is applied, and inner boundary for radius is kept inside the event horizon, $r=1a$, thus allowing the wave propagate freely inside the horizon. Reflection coefficient is calculated at sufficiently far away from the horizon, and inside the horizon is dismissed from the calculation vie excision technique. If we compare the reflection calculated in the time domain for a given initial wave with the one in the frequency domain with given asymptotic solutions, results do not differ. Figure \ref{both} shows the comparison between two solution methods.

\section{Discussion}
\label{Discussion}
Superradiance phenomena is the analog of the Penrose process for rotating black holes. Energy extraction from the black hole analogy, i.e. the vortex defined in BEC is shown to be possible by examining the scattering process. Acoustic superradiance defined as amplification of the reflection coefficient to values greater then one.

 Although the superradiant experiments focuses on the shallow water wave, water vortex or optical systems rather than 2D-BEC vortex, theoretically shown that it is possible and due to quantum nature it may be a great candidate for other phenomenas like Hawking Radiation \cite{richartz2015rotating},\cite{steinhauer2016observation},\cite{PhysRevLett.106.021302}.

In this work, we investigated the amplified scattering of acoustic waves propagating in a BEC, from a vortex state  with a constant background density, by obtaining both time-domain and asymptotic frequency domain solutions numerically. Time-domain study amounts for solving the Klein-Gordon equation which governs the propagation of sound waves in the presence of vortex in an analogy to scalar field propagation in curved space-time of a black-hole. It is worth to note that the classical (macroscopic) wave function of the BEC represents the classical space-time of General Relativity only when probed at long-enough wavelengths such that it behaves purely hydrodynamically. The major outcome of the study is to demonstrate a good spectral agreement of the superradiance (reflection coefficient) as obtained from full time-domain calculations and from the asymptotic frequency domain calculations. This strengthens the validity of the spectral analysis based only to the asymptotic solutions, which can be calculated with significantly less computational resource compared to that required by the time-domain calculations.
The frequency spectrum analysis gives further insight to the superradiance condition that is given in terms of the modulation frequency of the incident wave and the angular speed of the vortex as $\omega<m\Omega$. The maximum superradiance shows a gradual increase with increasing $\Omega$. For a given $\Omega$, superradiance is maximized for $\omega/\Omega \approx 0.68-0.69 $ when $\Omega<1$ and for $\omega/\Omega \approx 0.75-0.85 $ when $\Omega>1$.  Typically, a strongly modulated Gaussian pulse is able to acquire more energy through scattering.
As a final note, the theoretical and computational formulation presented in this work is suitable for the implementation of different background density profiles, which can potentially extend the exploration beyond the constant background-density approximation. This will be pursued in subsequent studies.

\section*{Acknowledgement}
 Bet\"ul Demirkaya is supported by TUB\.{I}TAK-B\.{I}DEB 2211 National Scholarship Program for PhD Students.

\bibliographystyle{unsrt}
\bibliography{bibfile2}

\end{document}